\def\b{\beta}\def\d{\delta}\def\e{\epsilon}
\def\h{\theta}
\def\k{\kappa}\def\l{\lambda}\def\m{\mu}\def\n{\nu}\def
\p{\pi}\def\s{\sigma}
\def\y{\eta}

\def\D{\Delta}

\def\V{\varphi}

\def\de{\partial}
\def\id{\equiv}\def\ha{{1\over 2}}

\def\({\left(}\def\){\right)}\def\[{\left[}\def\]{\right]}
\def\bdot{\!\cdot\!}

\def\mn{{\mu\nu}}

\def\coo{coordinates }

\def\pb{Poisson brackets }

\def\sch{Schwarzschild }
\def\poi{Poincar\'e }

\def \schr{Schr\"odinger }

\def\wrt{with respect to }\def\ie{i.e.\ }
\def\eom{equations of motion }
\def\cor{commutation relations }
\def\ur{uncertainty relation }

\def\kp{$\k$-\poi }

\def\section#1{\bigskip\noindent{\bf#1}\smallskip}

\def\PL#1{Phys.\ Lett.\ {\bf#1}}
\def\PRL#1{Phys.\ Rev.\ Lett.\ {\bf#1}}
\def\PR#1{Phys.\ Rev.\ {\bf#1}}\def\CQG#1{Class.\ Quantum Grav.\ {\bf#1}}

\def\PRS#1{Proc.\ R. Soc.\ Lond.\ {\bf#1}}
 \def\IJMP#1{Int.\ J. Mod.\ Phys.\ {\bf #1}}
\def\MPL#1{Mod.\ Phys.\ Lett.\ {\bf #1}}

\def\ref#1{\medskip\everypar={\hangindent 2\parindent}#1}
\def\beginref{\begingroup
\bigskip
\centerline{\bf References}
\nobreak\noindent}
\def\endref{\par\endgroup}

\def\bx{{\bf x}}\def\bp{{\bf p}}\def\bL{{\bf L}}

{\nopagenumbers
\line{}
\vskip30pt
\centerline{\bf Noncommutative geometry of the quantum clock}

\vskip60pt
\centerline{
{\bf S. Mignemi}$^{1,2}$\footnote{$^\ddagger$}{e-mail: smignemi@unica.it}
and {\bf N. Uras}$^{1}$}

\vskip10pt

\smallskip
\centerline{$^1$Dipartimento di Matematica e Informatica, Universit\`a di Cagliari}
\centerline{viale Merello 92, 09123 Cagliari, Italy}
\smallskip
\centerline{$^2$INFN, Sezione di Cagliari, Cittadella Universitaria, 09042 Monserrato, Italy}

\vskip60pt

\centerline{\bf Abstract}
\medskip
{\noindent
We introduce a model of noncommutative geometry that gives rise to the uncertainty relations recently derived from
the discussion of a quantum clock. We investigate the dynamics of a free particle in this model from the point of
view of doubly special relativity and discuss the geodesic motion in a \sch background.}
\vskip10pt
{\noindent

}
\vfill\eject\

}

Several proposal exist for modifications of the Heisenberg uncertainty relations when effects of gravity are taken into account [1].
These are usually based on thought experiments and involve a dimensional parameter of the order of the Planck length (or mass) that
sets the scale of the deformation.
If one assumes that the Heisenberg algebra is deformed, such modifications can of course be derived formally by standard quantum
mechanical arguments.

Deformed Heisenberg algebras have been considered in the literature mainly in relation with theories involving deformations of the
Lorentz symmetry, like doubly special relativity (DSR) [2] or noncommutative geometries, especially of the \kp class [3].
In fact, these theories are strongly related, although DSR investigates the deformations mainly from a classical (\ie non-quantum)
point of view.
In these theories, the deformations are due to the introduction of a new fundamental scale, that cannot be invariant under the
standard Lorentz transformations and whose appearance is justified as an effect of quantum gravity. For example, \kp models are
based on the deformed \cor of space and time \coo $[x_0,x_i]=ix_i/\k$, with $\k$ a constant proportional to the Planck mass\footnote{$^1$}
{We adopt the signature $(-1,1,1,1)$ and denote spacetime \coo as $(x_0,x_i)=(ct,{\bf x})$.}, which imply a deformation of the full
Heisenberg algebra. Note that such deformation is not unique and different models (usually called bases of the \kp algebra) can be
defined, leading to different modifications of the uncertainty relations.

While the modifications of the uncertainty relations considered in the literature usually concern the position-momentum relations,
in a recent paper [4] a thought experiment has been discussed, which predicts an uncertainty relation
connecting the measure of time and spatial intervals, given by
$$\D r\D t\ge\b,\eqno(1)$$
where $r=\sqrt{{\bf x}^2}$ is a radial coordinate and $t$ is time. The constant $\b$ is given in terms of the Planck length $L_P$
by $\b=L_P^2/c$. Clearly, this uncertainty relation can be interpreted as due to noncommutativity of spatial and time
coordinates, in analogy with the \kp model cited above.
The thought experiment is based on an ideal "quantum clock", namely a device that measures time by counting the
decays of a sample of radioactive matter, that was first devised in ref.\ [5].

The quantum clock is defined as follows:
given a set of $N$ radioactive particles of mass $m$, with total mass $M=Nm$,
the mean number of decays in a time interval $\D t$ is $\D N=\l N\D t$, with variance $\s_N=\sqrt{\l N\D t}$. Therefore, it
is possible to measure a time interval counting the number of decays.
The relative error $\e$ in the time measurement will be
$$\e={\s_t\over\D t}={1\over\sqrt{\l N\D t}},\eqno(2)$$
where $\s_t=\s_N/\l N$.
In order to measure short time intervals with small relative error it is therefore necessary to increase $N$.
From eq.\ (2) it follows that
$$\D t={1\over\e^2\l N}={m\over\e^2\l M},\eqno(3)$$
or, in terms of the rest energy of the particles $E=mc^2$,
$$\D t={E\over\e^2\l Mc^2}.\eqno(4)$$

Now, from the Heisenberg uncertainty relation, one has for each particle
$$\d E\,\d t\ge\hbar/2,\eqno(5)$$
where $\d E$ and $\d t$ are the uncertainties in the energy and time measurements. But $\d E<E$, $\d t<1/\l$, and hence
$${E\over\l}\ge\hbar/2.\eqno(6)$$
Using (4), one finally obtains
$$\D t\ge{\hbar\over2\e^2c^2M},\eqno(7)$$
which gives a lower limit for the mass of a clock capable of measuring time intervals with accuracy $\D t$.

However, it is not possible to arbitrarily increase the mass of the clock holding it in a small volume, since
the radial size $R$ of the clock must be such that a black hole cannot form, and therefore greater than its \sch radius,
$$R>{2GM\over c^2}.\eqno(8)$$
Setting $\D r=R$, with $r=\sqrt{{\bf x}^2}$, from (7) it follows that
$$\D r\D t\ge{G\hbar\over c^4},\eqno(9)$$
which is the relation (1). In a quantum theory, this uncertainty relation can be derived  assuming (up to numerical factors)
the commutation relation $[t,r]=i\b$.
In fact, by the usual quantum mechanical argument, in the case of vanishing expectation values of $r$ and $t$:
$$\D r\D t\ge\ \ha\,\big|\langle\,[r,t]\,\rangle\big|={\b\over2}.\eqno(10)$$

It is therefore natural to assume that the \ur (1) can be obtained starting from a deformation of the Heisenberg algebra of the
kind investigated in  noncommutative geometry or in DSR theories.
In particular, a deformed commutation relation leading to (1) is, in relativistic notation\footnote{$^2$}{In the following we
use natural unities, $\hbar=c=G=1$.},
$$[x_0,x_i]=i\b{x_i\over r},\eqno(11)$$
which clearly implies a noncommutative geometry, and in particular recalls the \kp \cor $[x_0,x_i]=ix_i/\k$.
It is therefore likely that it can be obtained from a similar construction.

Actually, assuming (11), one can construct several deformations of the Heisenberg algebra obeying the Jacobi identities.
We consider here the simplest deformation compatible with (11), and investigate its classical limit, with commutators replaced by
Poisson brackets, and its DSR implementation.
Investigation of the quantum theory may result difficult, since the \cor (11) are nonlinear in the \coo $x_i$, contrary to the
models usually investigated in the context of noncommutative geometry.

We define the deformed algebra through the \pb
$$\eqalignno{&\{x_i,x_j\}=0,\qquad\{x_0,x_i\}=\b{x_i\over r},\qquad\{p_\m,p_\n\}=0,\qquad\{x_i,p_j\}=\d_{ij},&\cr
&\{x_i,p_0\}=0,\qquad\{x_0,p_0\}=-1,
\qquad\{x_0,p_i\}=-{\b\over r}\(p_i-{\bx\bdot\bp\over r^2}x_i\).&(12)}$$
It is easy to check that this algebra implies $\{x_0,r\}=\b$, as required, and that the \pb are covariant under spatial rotations.
This algebra can be realized in terms of canonical coordinates $\tilde x_\m$, $p_\m$ by the simple  rule
$$x_0=\tilde x_0-\b{\tilde x_ip_i\over r},\qquad x_i=\tilde x_i,\eqno(13)$$
while the momenta maintain their canonical form.

The \pb (12) cannot however be covariant under boosts, since $x_0$ and $r$ are not. Defining the Lorentz generators as
$J_\mn\id\tilde x_\m p_\n-\tilde x_\n p_\m$, so that the Lorentz algebra is not deformed,
the infinitesimal action on the spacetime \coo of a boost in the $i$ direction is given by $\d_L x_\m=\{J_{0i},x_\m\}$.
The covariance under boosts is obtained if their action is deformed so that
$$\d_L x_j=\d_{ij}\(x_0+{\bx\bdot\bp\over r}\),
\qquad\d_L x_0=-x_i-{\b\over r}\(p_0+{\bx\bdot\bp\over r^2}\)x_i+{\b\over r}\(x_0+\b{\bx\bdot\bp\over r^2}\)p_i.
\eqno(14)$$
The deformation of the action of boosts is a well-known consequence of the modification of the Heisenberg algebra.

Contrary to standard DSR models, the transformation rules of the momenta are instead not modified, and hence the \poi algebra is preserved.
It follows in particular that the Casimir invariant of the \poi algebra is $p^2$, as in special relativity.
One can therefore take as Hamiltonian for a free particle
$$H={p^2\over2m}.\eqno(15)$$
Starting from this Hamiltonian, and taking into account the deformed \pb (12), the Hamilton equations for a free particle are then
$$m\dot x_i=\{x_i,H\}=p_i,\qquad m\dot x_0=\{x_0,H\}=-p_0-{\b\over r}\(\bp^2-{(\bx\bdot\bp)^2\over r^2}\),\qquad\dot p_\m=\{p_\m,H\}=0,
\eqno(16)$$
where a dot denotes a derivative with respect to the evolution parameter.

The \eom can also be obtained varying the action
$$S=-\int ds\(x^\m\dot p_\m+\b{\bx\bdot\bp\over r}\dot p_0+H\).\eqno(17)$$
In fact, varying \wrt  $x^\m$ and $p_\m$, one gets
$$\dot x_0=-{p_0\over m}-\b\({\dot\bx\bdot\bp+\bx\bdot\dot\bp\over r}-{\bx\bdot\bp\ \bx\bdot\dot\bx\over r^2}\),\qquad
m\dot x_i={p_i\over m}+\b{x_i\dot p_0\over r},$$
$$\dot p_0=0,\qquad\dot p_i+\b{p_i\over r}\dot p_0=0,\eqno(18)$$
which are equivalent to (16).

As usual in theories containing deformations of the Lorentz symmetry, some problems arise in the definition of the velocity [6].
In fact, in relativistic theories the 3-velocity of a particle can be defined either as $v^H_i={\de p^0\over\de p^i}$ or as
$v^K_i={\dot x_i\over\dot x_0}$, and the two definitions can yield different results.
In our model, the first definition gives the standard relativistic relation $v^H_i=p_i/p_0$, while the second definition leads to
$$v^K_i={p_i\over p_0+{\b\over r}\(\bp^2-{(\bx\cdot\bp)^2\over r^2}\)}={p_i\over p_0+\b{\bL^2\over r^3}},\eqno(19)$$
where $\bL=\bx\times\bp$ is the angular momentum.
%This is in contrast with DSR models, where usually is the velocity $v^H$ that is deformed. This fact can be ascribed to the fact that
%in the present case the action of the Lorentz group is deformed only on spacetime.
The velocity $v^K_i$ is always less than 1 for positive $\b$.

Notice that this result is in contrast with most DSR models where $v^K_i$ has the standard relativistic form, while $v^H_i$ is deformed.
This is related to the fact that in our case the \poi algebra, and hence the dispersion relation for particles, is not deformed, but its
action on \coo is.

In this respect, our model differs from the standard DSR models, that assume a deformation of the action of the Lorentz group on
momentum space, rather than spacetime.
However, like in that case, the deformation extends to the full phase space, since the position-momentum \cor are deformed (cf.\ (12)).
This implies also a deformation of the standard relativistic position-momentum uncertainty relations.

The phenomenological implications of our model are not easy to disclose.
The problem is that we have chosen a minimal deformation compatible with (11), affecting only the time variable,
and hence its effects can appear only in fully relativistic situations.

The simplest effect in DSR phenomenology is the time delay in the detection of photons of different energies coming from a distant source.
In [7] it has been shown  that, taking into account the effect of the nontrivial action of translations, the time delay predictions calculated
using either the phase or the group velocity are identical at least for some models of DSR.
In our case  no time delay is present. In fact, for particles moving on a straight line from the source to the observer, both phase
and group velocity are equal and coincide with the relativistic ones, since in that case $\bL=0$.
Some effects could however be present if the particle travels on a curved trajectory.

As an illustration, we therefore consider the orbital motion in \sch spacetime, although it is not likely that the associated corrections be
observable in practice.
We shall show that deviations from the predictions of relativity only occur in the time of travel, while the trajectories are
unaltered.

In fact, due to the conservation of angular momentum, the problem can as usual be reduced to 1+2 dimensions. Going to spherical \coo
$$t=x_0=-x^0,\qquad r=\sqrt{(x^1)^2+(x^2)^2},\qquad\h=\arctan{x^2\over x^1},\eqno(20)$$
with momentum components
$$p_t=p_0,\qquad p_r={x^1p_1+x^2p_2\over\sqrt{(x^1)^2+(x^2)^2}}={\bx\cdot\bp\over r},
\qquad p_\h\id J_{12}=x_1p_2-x_2p_1,\eqno(21)$$
it is easy to check that the only nontrivial brackets are
$$\{t,r\}=\b,\qquad\{t,p_t\}=-1,\qquad\{r,p_r\}=1,\qquad\{\h,p_\h\}=1.\eqno(22)$$

The Hamiltonian for the motion of a free particle of mass $m$ in \sch spacetime is\footnote{$^3$}
{We use this unusual normalization in order to keep track of possible breakdowns of the equivalence principle.}
$$H={1\over2m}\[-{p_t^2\over A}+Ap_r^2+{p_\h^2\over r^2}\],\eqno(23)$$
with
$$A(r)=1-{2M\over r}.\eqno(24)$$
The Hamilton equations read
$$m\dot t={p_t\over A}+{\b M\over r^2}\(p_r^2+{p_t^2\over A^2}\)-\b{p_\h^2\over r^3},
\qquad m\dot r=Ap_r,\qquad m\dot\h={p_\h\over r^2},\eqno(25)$$
$$\dot p_t=\dot p_\h=0,\qquad m\dot p_r=-{M\over r^2}\(p_r^2+{p_t^2\over A^2}\)+{p_\h^2\over r^3}.\eqno(26)$$
Two conserved quantities are present,
$$p_t\id mE=mA\dot t-{\b MA\over r^2}\(p_r^2+{p_t^2\over A^2}\)+\b{p_\h^2A\over r^3},\qquad p_\h\id ml=mr^2\dot\h,\eqno(27)$$
where we have introduced the normalized momenta $E$ and $l$.
Moreover, $p_r$ can be obtained in terms of the other momenta from the constraint $H=-m/2$, as
$$p_r^2={p_t^2\over A^2}-{p_\h^2\over Ar^2}-{m^2\over A}.\eqno(28)$$

From (25) it follows that the equation of the orbits is independent of $\b$ and has the standard relativistic form
$${dr\over d\h}={r^2Ap_r\over p_\h}.\eqno(29)$$
The solution of (29) can be obtained in the usual way by an expansion in the parameter $\y={M^2 \over l^2}$, as [8]
$$u\id{1\over r}\sim{M\over l^2}(1+e\cos\h'),\eqno(30)$$
where $e$ is the eccentricity of the orbit, related to the energy $E$ by
$$E^2\sim1-\y(1-e^2),\eqno(31)$$
and [8]
$$\h'\sim\h-\y(3\h+e\sin\h),\eqno(32)$$
from which the standard perihelion shift $\D\h=6\p\y$ follows.

However, the time dependence of the orbit is modified, and so its period. In fact, from (25) and (28),
$${dt\over d\h}={E\over lAu^2}+\b m\[{M\over l}\({2E^2\over A^2}-{l^2u^2+1\over A}\)-lu\],\eqno(33)$$
and after substituting (30), (31) and (32) one obtains, up to order $\y$,
$${dt\over d\h'}\sim{l^3\over M^2}\ {1+3\y(1+e\cos\h')\over(1+e\cos\h')^2}+{\b mM\over l}
\[-e\cos\h'+\y(3+2e^2+e\cos\h'-e^2\cos^2\h')\].\eqno(34)$$

To compute the period $T$ of the orbit (defined as the time between two successive passages through the perihelion)
we integrate (34) in $\h'$ between 0 and $2\p$. We get
$$T={2\p l^3[1+3\y(1-e^2)]\over M^2(1-e^2)^{3/2}}+{2\p\b mM^3(3+e^2)\over l^3}.\eqno(35)$$
The first term is the classical one [8], while the second comes from the deformation of the symplectic structure.
At leading order, the relative correction to the orbital period due to the
second term is of order ${\b M^5\over l^6}\,m$, and depends linearly on the mass of the planet.
Thus this correction breaks the equivalence principle, which is a common feature of DSR models [9].

The relative correction for the orbital period of the Earth would be  $10^{-25}$, \ie of the order of $10^{-20}$
seconds, and similarly for other planets. The corrections are therefore extremely tiny and there is no chance to detect them.
A different system that might give rise to stronger effects is a particle orbitating in a cyclotron. Also in this case the
frequency should depend on the energy of the particle. The problem is presently being investigated.

It would also be interesting to consider the effects of our deformation in the quantum domain.
First of all, we notice that, although the position-momentum \pb in (11) look odd, they take a perfectly standard form if
spherical spatial coordinates $r$, $\h$, $\V$ are used, namely
$$\{t,p_t\}=-1,\qquad\{r,p_r\}=\{\h,p_\h\}=\{\V,p_\V\}=1.\eqno(36)$$
Thus, under quantization in these coordinates, only the $t$-$r$ commutation relations are deformed and hence the relativistic
Heisenberg uncertainty relations stay unchanged, except (1).

One may ask if any effect might nevertheless occur in the relativistic hydrogen atom.
To study this, one should find a Hilbert space realization of the \cor corresponding to the \pb (22).
It is easy to see that the only difference from the standard realization is in the time operator, which now reads
$t\to t-i\de/\de r$, cfr.\ (13). However in the relativistic \schr (or Dirac) equation for the hydrogen atom
only the operators $p_\m$ and $r$ appear and therefore no corrections arise in the energy spectrum.
The only possible effects could occur for time-dependent observables.

To conclude, it  seems that our model predicts only extremely tiny observable effects, related to the measure of time.
However, we again remark that different models compatible with (11) could be constructed, presenting a more involved algebra
than (12), that includes deformations also in the momentum sector, and is therefore more similar to standard DSR theories.
The aim of our investigation has been anyway to investigate the deformations of the Heisenberg algebra compatible with the
quantum clock uncertainty relations presenting a minimal departure from general relativity, rather than phenomenologically
relevant ones.

It would also be interesting to study our model from the point of view of noncommutative geometry, to obtain for
example the correct momentum addition law.
This problem is not trivial because, as mentioned above, the \cor (11) are nonlinear in the coordinates and this may give rise
to considerable technical problems.

\beginref
\ref [1] T. Padmanabhan, \CQG{3}, 911 (1986); D. Amati, M. Ciafaloni and G. Veneziano, \PL{B216}, 41 (1989);
M. Maggiore, \PL{B304}, 65 (1993); F. Scardigli, \PL{B452}, 39 (1999); R. Adler and D.J. Santiago, \MPL{A14}, 1371 (1999).
\ref [2] G. Amelino-Camelia, \PL{B510}, 255 (2001); \IJMP{D11}, 35 (2002);
J. Magueijo and L. Smolin, \PRL{88}, 190403 (2002); \PR{D67}, 044017 (2003).
\ref [3] J. Lukierski, H. Ruegg, A. Novicki and V.N. Tolstoi, \PL{B264}, 331 (1991);
J. Lukierski, A. Novicki and H. Ruegg, \PL{B293}, 344 (1992); S. Majid and H. Ruegg, \PL{B334}, 348 (1994).
\ref [4] L. Burderi, T. Di Salvo and R. Iaria, \PR{D93}, 064017 (2016).
\ref [5] H. Salecker and E.P. Wigner, \PR{109}, 571 (1958).
\ref [6] P. Kosi\'nski and P. Ma\'slanka, \PR{D68}, 067702 (2003); S. Mignemi, \PL{A316}, 173 (2003);
 M. Daskiewicz, K. Imilkowska, J. Kowalski-Glikman, \PL{A323}, 345 (2004).
\ref [7] G. Amelino-Camelia, N. Loret and G. Rosati, \PL{B700}, 150 (2011).
\ref [8] C. Darwin, \PRS{A263}, 39 (1961); P.A. Geisler and G.C. McVittie, AJ {\bf 10}, 14 (1965).
\ref [9] S. Mignemi and R. \v Strajn, \PR{D90}, 044019 (2014).

\endref
\end